# Beyond object identification: How train drivers evaluate the risk of collision

Romy Müller & Judith Schmidt

*Faculty of Psychology, Chair of Engineering Psychology and Applied Cognitive Research, TUD Dresden University of Technology, Dresden, Germany*

Corresponding author:

Romy Müller

Chair of Engineering Psychology and Applied Cognitive Research

TUD Dresden University of Technology

Helmholtzstraße 10, 01069 Dresden, Germany

Email: romy.mueller@tu-dresden.de

Phone: +49 351 46335330

ORCID: 0000-0003-4750-7952

**Abstract**

When trains collide with obstacles, the consequences are often severe. To assess how artificial intelligence might contribute to avoiding collisions, we need to understand how train drivers do it. What aspects of a situation do they consider when evaluating the risk of collision? In the present study, we assumed that train drivers do not only identify potential obstacles but interpret what they see in order to anticipate how the situation might unfold. However, to date it is unclear how exactly this is accomplished. Therefore, we assessed which cues train drivers use and what inferences they make. To this end, image-based expert interviews were conducted with 33 train drivers. Participants saw images with potential obstacles, rated the risk of collision, and explained their evaluation. Moreover, they were asked how the situation would need to change to decrease or increase collision risk. From their verbal reports, we extracted concepts about the potential obstacles, contexts, or consequences, and assigned these concepts to various categories (e.g., people's identity, location, movement, action, physical features, and mental states). The results revealed that especially for people, train drivers reason about their actions and mental states, and draw relations between concepts to make further inferences. These inferences systematically differ between situations. Our findings emphasise the need to understand train drivers' risk evaluation processes when aiming to enhance the safety of both human and automatic train operation.

# 1 Introduction

Although trains have become a relatively safe means of transportation, collisions keep occurring. Their types and frequencies may vary between locations, but they remain a major concern around the world (Ballay et al., 2022; Hampel et al., 2023; Rosić et al., 2022; Skládaná et al., 2016; Zhang et al., 2023). Given the mass and speed of trains, collisions can have disastrous consequences for the people or objects a train is colliding with, the train, the driver, the passengers, and sometimes even for the infrastructure or third parties not directly involved in the accident. Collisions with people are particularly concerning due to the combination of their frequency and severity: people illegally enter railway territory on a regular basis (Skládaná et al., 2016) and when they get hit by a train, this usually results in fatalities (Hampel et al., 2023). In the aftermath of such accidents, train drivers have to deal with acute psychological disturbances (Limosin et al., 2006).

Current endeavours to enhance railway safety envision a use of artificial intelligence (AI) to partly or fully automate train operation (ATO). These approaches focus on sensing and identifying potential obstacles to avoid collisions. However, this focus seems overly limited. Imagine a train entering a station. An AI system might classify the objects on the platform as people with high certainty. Does this make the situation dangerous? It depends. In fact, it is normal that people are standing there – this is what platforms are made for. The relevant question is what they are doing and whether they might get in conflict with the train. In more abstract terms, potential obstacles are omnipresent. To arrive at a valid risk evaluation, it is necessary to infer whether a potential obstacle is likely to become an actual one. Thus, false positives need to be eliminated in order to single out true positives. A challenge is that false and true positives often belong to the same object class. For instance, normal passengers waiting for their train are people in the vicinity of the tracks, just like the infrequent outliers who might enter the danger zone by accident, as a result of a criminal act, or in an attempt to end their lives. Due to a train's long braking distances, there usually is insufficient time to simply wait until the situation gets disambiguated as the train is approaching. Instead, it is necessary to infer how the situation might develop before this development is manifested in concrete, observable events.

Given this challenge, the limited focus of contemporary AI research on object identification seems puzzling. In the psychological literature, it has been known for decades that situation awareness requires more than mere identification: to understand dynamic situations, agents (either human or automated) need to interpret what is going on at the moment in order to predict how situations are likely to unfold in the future. In the scenario above, it is necessary to pick up visual cues (e.g., about a person's appearance and behaviour as well as about context factors such as time of day) in order to interpret what is going on (e.g., what the person might be intending) and anticipate how the situation is likely to unfold (e.g., whether the person might enter the danger zone). To help artificial agents establish situation awareness, it would be desirable if AI research could build on psychological knowledge about train drivers' strategies. Surprisingly, many studies with human train drivers have adopted the same limited focus as contemporary AI research: they mainly investigated basic scene perception and object identification. Although task analyses of train driving are aware of the need to interpret and anticipate events, they mainly relate these abilities to route knowledge without specifying how situation awareness can emerge from the visual cues available in a particular situation.

The present study contributes to a deeper understanding of how train drivers evaluate the risk of collision. This presupposes a detailed description of what they perceive and infer when processing potential obstacles. To this end, we conducted image-based expert interviews with 33 train drivers, investigating what concepts they use to reason about collisions. Before presenting the details of the study, we will highlight the limitations of contemporary AI in avoiding collisions, summarise previous research on train drivers' perception and cognition, and explain how the focus of this research should be extended.

# 2 Theoretical background

## 2.1 Can collisions be avoided by means of AI?

*2.1.1 Promises and limitations of AI*

Enhancing the safety and efficiency of train operation is an ongoing challenge. In recent years, concepts of AI-based automatic train operation have gained popularity (Flammini et al., 2022). An important sub-challenge of this endeavour is to avoid collisions or reduce the harm resulting from them. To this end, current research efforts mainly focus on obstacle detection and identification (for overviews see Cao et al., 2024; Ristić-Durrant et al., 2021). The respective AI systems rely on data from different sensors (e.g., camera, infrared, LiDAR), process these data with image classifiers based on deep neural networks to identify potential obstacles (e.g., person on the tracks), and then the results can be used to make operational decisions (e.g., braking). However, AI is not perfect and thus there are limits to AI-based collision prediction and avoidance.

Outside the railway domain, it is well-established that image classifiers can make bizarre mistakes (Firestone, 2020; Geirhos et al., 2020). For instance, they may fail to recognise a cow in an unusual context like a beach (Beery et al., 2018), they typically rely on texture more than shape and thus classify a cat with wrinkled skin as an elephant (Geirhos et al., 2019), and they can get derailed, quite literally, when small patches of pixels are replaced in such a way that humans would not even notice the change (Jacobsen et al., 2019). In the railway domain, such mistakes may be even more likely. This is due to a number of technical challenges. For instance, suitable datasets are scarce, classification performance deteriorates under difficult weather conditions, and small objects or unknown categories are hard to detect (Cao et al., 2024). Not surprisingly, such classification failures and technical issues have raised concerns about the safety of AI systems for collision avoidance (Rosić et al., 2022). Much of the corresponding discussion focuses on whether AI will be robust enough in identifying potential obstacles. For instance, can an AI system reliably detect a person near the tracks? Can it distinguish between a shadow and a tree obstructing the tracks?

Although these certainly are relevant questions, we doubt that inaccurate object identification is the problem we should be most concerned about.[1] Additional challenges of automatic train operation have been highlighted in studies identifying the hidden roles of train drivers (Jansson, Fröidh, et al., 2023; Jansson, Olsson, et al., 2023; Karvonen et al., 2011). For instance, train drivers need to take care of passengers and flexibly act in exceptional situations. But even if you just focus on driving, and within this task you just single out the avoidance of obstacles – is their identification the main challenge? We ought to be sceptical about this as the world is full of distractors: things that could pose a threat in principle, but then upon closer inspection turn out not to. If emergency braking was initiated in all of these cases, train operation would get completely inefficient (cf. Rosić et al., 2022). A key problem with distractors is that they often belong to the same object classes as actual threats, but behave in a non-threatening way. For instance, people close to the tracks are normal in many contexts. Thus, AI systems would need to decide whether a potential obstacle really poses a risk and requires a response. How do humans handle such ambiguity?

---

[1] Some sources differentiate between object detection, recognition, and identification: detection means deciding whether an object is present (i.e., distinguishing it from the background), recognition requires determining the object class (e.g., human, animal, vehicle), and identification refers to a detailed description of the object (e.g., worker with a safety vest). We argue that all of these capabilities might be insufficient. In the remaining article, we will use the term "object identification" to refer to the full range of capabilities.

*2.1.2 How humans understand complex, dynamic situations*

Understanding complex, dynamic situations has been a subject of psychological research for decades. The concept of situation awareness (Endsley, 1995, 2017) assumes that people grasp dynamic situations on three levels. Perceiving the elements of the current situation only is a first step, which forms the basis for comprehension and projection. That is, people do not only identify what is currently present, but interpret what they see and anticipate how the situation is likely to unfold. This seems particularly relevant in the context of train operation due to the long braking distances and the resulting need to plan ahead (cf. Roth & Multer, 2009). Endsley's three-level model of situation awareness is just one of many psychological theories describing how humans handle complex, dynamic situations. Others, like Neisser's perceptual cycle (Neisser, 1976), Klein's recognition-primed decision making (Klein, 1989) or Rasmussen's decision ladder (Rasmussen, 1986) were proposed long ago and still shape contemporary research (Flach, 2015; Klein, 2008; Lintern, 2010; Plant & Stanton, 2015). They all share the assumption that humans do not merely identify objects but process this low-level information in combination with contextual constraints and prior knowledge to draw inferences. Given that these higher-order processes are common sense in psychology, one might assume that there are plenty of insightful studies spelling out how train drivers process potential obstacles, and their findings could readily be used to inspire AI research. Well, unfortunately not. In the following sections, we will review what is known about train drivers' visual processing strategies, and what is missing to understand how they evaluate the risk of collision.

## 2.2 How do train drivers process their visual environment?

*2.2.1 Train drivers in psychological studies*

One approach of understanding how train drivers process their environment is to monitor their actual behaviour. Train driving imposes high visual demands on drivers (Naweed & Balakrishnan, 2014) and the visual sense is their most important source of information to detect abnormalities (Jansson, Fröidh, et al., 2023). Thus, it is not surprising that numerous studies have tracked train drivers' eye movements, either in simulators or out on the tracks (Du et al., 2022; Guo et al., 2015; Itoh et al., 2000; Luke et al., 2006; Ma et al., 2024; Merat et al., 2002; Sun et al., 2019; Suzuki et al., 2019). These eye movements were found to depend on operational and personal factors. For instance, they differed between urban and rural environments (Guo et al., 2015; Luke et al., 2006) or between high and low speeds (Guo et al., 2015; Itoh et al., 2000; Suzuki et al., 2019). Moreover, they were affected by train drivers' overall job experience (Du et al., 2022; Sun et al., 2019), their experience with collisions (Ma et al., 2024), and their route knowledge (Itoh et al., 2000). A few studies have zoomed in on the detection of abnormal objects (Itoh et al., 2000; Ma et al., 2024; Suzuki et al., 2019). In these studies, detection performance was measured by having train drivers press a button or hit the brakes as soon as they notice an object. Detection was found to depend on factors such as the speed of the train, the curvature of the tracks, and train drivers' collision experience.

There is a critical limitation common to almost all behavioural studies investigating train drivers' visual strategies: they did not go beyond the processing of low-level aspects like location and identity. In that sense, they are quite similar to contemporary AI research. This limitation was addressed in cognitive task analyses that usually relied on interviews with experienced train drivers (e.g., Hamilton & Clarke, 2005; McLeod et al., 2005; Naweed, 2014; Rose & Bearman, 2012; Roth & Multer, 2009; Zoer et al., 2014). Given that these studies aimed to understand the activity of train driving as a whole, they suffered from limitations opposite to those of behavioural studies. While behavioural studies zoomed in on a low-level aspect of train drivers' visual processing, task analyses tended to describe it on a very high level of abstraction (e.g., "perceptual processing", "cognitive processing") without specifying the underlying mechanisms. Perhaps the most detailed analysis is the one by Roth and Multer (2009). Their task model shows considerable resemblance with well-known models of situation awareness (e.g., Endsley, 1995; Neisser, 1976). It is beyond the scope of this article to review any particular task analysis. Instead, we will summarise current knowledge about the three levels of situation awareness in the context of train driving: What visual cues do train drivers use, how do they interpret these cues, and how do they anticipate future events?

*2.2.2 What visual cues do train drivers use?*

When describing their work in challenging situations, most train drivers emphasise the need to manage the density of information coming in from the environment (Naweed & Balakrishnan, 2014). Some studies have extracted the visual cues train drivers pick up during normal operation (Luke et al., 2006; Roth & Multer, 2009). First, train drivers monitor specific objects and events. That is, they attend to the locations of agents (e.g., other trains, workers, trespassers) and other potential hazards. Aside from that, they consider the broader context. Thus, a second type of cue refers to environmental constraints such as the weather and sight-related factors (e.g., fog, darkness). Third, train drivers keep track of the infrastructure, considering both fixed physical elements (e.g., bridges, tunnels, stations) and dynamic physical elements (e.g., construction sites). Fourth, they are constantly aware of relevant operational aspects (e.g., their current location and speed, signals, train characteristics).

Two studies specifically focused on potential collisions and described what cues train drivers use in these situations. First, Rosić et al. (2022) interviewed 68 train drivers to assess the type and frequency of different obstacles. In their discussion, they briefly mentioned which features of these obstacles were used by train drivers to evaluate the risk of collision. Such cues included the obstacle's distance, position, movement, and physical features. The latter was illustrated by stating that the risk associated with a cardboard box depended on its size, position, and content. Furthermore, train drivers considered the type of train as it affects the impact of collisions. For people, it was mentioned that train drivers attended to their body language Unfortunately, these cues were not reported in a systematic manner.

In another study that reported cues used by train drivers to reason about potential collisions, Tichon (2007) conducted focus group interviews using the critical decision method (an interview technique suitable for eliciting the details of exceptional situations). However, only one to six cues per incident were reported, and often they simply described the basic type of situation (e.g., "children crossing tracks before tunnel entrance"). Interestingly, several cues did not only identify potential obstacles but included the broader context. In particular, train drivers considered their current activity (e.g., approaching or leaving a station) as well as the signals and warnings they received (e.g., encountering railway detonators, warning via signals, receipt of information about track conditions). Taken together, it seems like train drivers use a variety of cues beyond the mere identity of potential obstacles. However, previous studies did not provide sufficiently detailed accounts of these cues.

*2.2.3 How do train drivers interpret the cues they pick up?*

The literature base is thin regarding the cues train drivers use, but it is even thinner regarding their interpretations. Based on the available studies, one might conclude that once an object or event is detected, the situation is clear – after all, previous studies typically did not discuss how train drivers infer whether the things they see actually pose a threat. However, there are a few exceptions.

First, in interviews with metro drivers, Karvonen et al. (2011) determined train drivers need to interpret and evaluate risks, for instance when a person at a station looks abnormal. Unfortunately, the authors just mentioned such interpretations anecdotally but did not investigate how drivers make them. Still, their anecdotal evidence is interesting, for instance when they listed examples of people who need to be monitored with special care. This need for enhanced monitoring depended on people's identity (e.g., children), behaviour (e.g., standing in the danger zone), interaction with objects (e.g., skateboards, wheelchairs), and on the inferences drivers derived from these visual cues (e.g., visually impaired, drunk, acting in a threatening way). These examples suggest that drivers go well beyond simply identifying an obstacle but interpret its behaviour and appearance to draw further conclusions.

A second exception stems from the study mentioned above that assessed collision frequencies (Rosić et al., 2022). In their discussion, the authors provided examples for inferences train drivers derived from a potential obstacle's identity. For instance, in case of track workers, train drivers inferred that these are professionals that are trained and medically fit to move out of the way. Accordingly, workers may be closer to the tracks than ordinary people, without train drivers getting overly concerned. Conversely, children or people with special needs are inferred to pose a higher risk.

Finally, some studies argued that train drivers infer people's mental states, for instance when considering what people can see at level crossings (Roth & Multer, 2009). Taken together, while the available evidence suggests that train drivers interpret situations, previous studies only reported such interpretations in the form of examples.

*2.2.4 How do train drivers anticipate future events?*

The available evidence also suggests that train drivers actively anticipate how a situation might unfold. For instance, they generate anticipations of movement authority, infrastructure, and human behaviour (Naghiyev et al., 2016). Although numerous studies stressed the importance of anticipation (Karvonen et al., 2011; McLeod et al., 2005; Naghiyev et al., 2016; Naweed, 2014; Phillips & Sagberg, 2014; Zoer et al., 2014), few of them specified what exactly these anticipations referred to and how they were generated. A notable exception is the task analysis by Roth and Multer (2009), which put a strong focus on looking ahead and planning. In fact, they concluded that the majority of train drivers' attention is focused on anticipating what is likely to happen next. Based on this anticipation, train drivers monitor their environment: knowing what to expect allows them to quickly direct their attention to events that are out of the ordinary (i.e., pattern matching).

In previous studies, it has not always been clear what train drivers' anticipations are based on. Some exclusively portrayed anticipations as a consequence of route knowledge (McLeod et al., 2005; Naghiyev et al., 2016), while others related it to the interplay between route knowledge and monitoring (Naweed, 2014; Roth & Multer, 2009). Only few explicitly linked anticipation to specific cues observed in the current situation (Tichon, 2007). For instance, based on railway detonators, train drivers predicted that workers are present further down the tracks. Based on children trespassing, train drivers predicted that they might have placed objects on the tracks, that they might throw stones, and that more children might be present.

Understanding how anticipations are generated from observable cues is important, because cue interactions can give rise to paradoxical effects (Rosić et al., 2022). For instance, a potential obstacle's position relative to the tracks is ambiguous as it interacts with movement direction and speed. Accordingly, train drivers may consider a vehicle directly on a level crossing as less dangerous than a vehicle further away but approaching at high speed. The former will be gone once the train arrives, while the latter might actually become an obstacle. Once again, only anecdotal evidence was reported in previous studies, and to date there is no systematic investigation of how train drivers perceive, interpret, and anticipate situations to evaluate the risk of collisions.

## 2.3 Present study

To infer how AI should assess the risk of collision, it is desirable to know how train drivers generate situation awareness. Similar to contemporary AI research, most behavioural studies of train driving adopted a narrow focus on eye movements and object detection. Their results tell us where train drivers look and what hazards they notice, but not how they use this information to evaluate the risk of collision. Conversely, most cognitive task analyses were quite broad in their focus, and few of them addressed potential collisions. Some described the cues train drivers use, and some even tackled interpretation and anticipation. However, these accounts were usually restricted to anecdotal evidence based on a few examples. In the present study, we aimed to better understand how train drivers evaluate the risk of collision.

To specify our research question, we first conducted a pilot study with six train drivers. In an unstructured interview, they described their experiences with obstacles (including collisions and near misses) and we asked a variety of questions about these situations. In line with previous research, we found that uncertainty and anticipation played a major role. To enable anticipation, train drivers drew inferences from features of the obstacle and context, while also considering the interactions between these cues. Moreover, the pilot study revealed another major source of anticipation that had not been as prominent in previous research: the presumed intentions of people, which train drivers inferred from various observable cues such as a person's identity or body language.

In the main study, we specifically investigated what cues train drivers use and how they integrate them. Based on the pilot study and the available literature, we expected train drivers to rely on higher-level inferences when evaluating the risk of collision. That is, we hypothesised that they would not only use features that are in the focus of contemporary AI research, such as an obstacle's identity, its location, or its movement. Rather, and especially when evaluating the risk associated with people, we expected train drivers to rely on non-observable concepts. This might be abstract descriptions of people's actions (i.e., behaviour with meaning), inferences about their mental states (e.g., perception, knowledge, intentions), or anticipations of future events (e.g., entering the tracks). In sum, we expected train drivers to go beyond what is currently visible when evaluating the risk of collision.

When speaking about inferences, we need to acknowledge that most of experts' risk evaluation proceeds implicitly (Klein, 1989, 2008): based on a rapid recognition of familiar patterns, they intuitively categorise situations, instead of pondering about the contributing factors and their relations. Therefore, we needed a knowledge elicitation technique that relies on concrete situations instead of abstract reasoning. At the same time, these situations should not provide too many cues about their dynamics in order to keep experts from making a quick, implicit judgment that they may not be able to verbally explain. Moreover, we wanted to gain more general insights instead of only describing particular situations in detail. In short, we needed a method that is clearly situated, but also allows for explicit analysis and counterfactual reasoning. To this end, we conducted expert interviews with train drivers based on images of potential collision situations (e.g., passengers at train stations, workers close to the tracks, trees leaning towards or obstructing the tracks). Drivers were asked to evaluate how dangerous each situation was, explain why, and describe how the situation would need to change in order to become less or more dangerous. We intentionally used static images that do not provide sufficient information about movement – neither of the train, nor of the potential obstacles. This was done to increase the likelihood of train drivers reporting these dynamic cues (or their absence), instead of simply perceiving them in a video, taking them for granted, and then basing their evaluations on this implicit information.

# 3 Methods

## 3.1 Data availability

All stimuli, instructions, transcripts, coding tables, relation graphs, aggregated data, and syntax files are made available via the Open Science Framework: https://osf.io/fcwgj/

## 3.2 Participants

Thirty-tree train drivers took part in the study. They were recruited via the Union of German Train Drivers (GDL), an online forum, the TUD Dresden University of Technology's participant pool, and personal contacts. All participants were male and had a job experience of 1 to 45 years ($M$ = 19.6, $SD$ = 13.1). They were currently driving different types of train (17 suburban and regional transport, 6 long distance transport, 10 freight) and most had driven more than one type of train in the past. Ten train drivers additionally worked as trainers. Five had already participated in the pilot study. A monetary compensation of 15€ per hour was offered. However, only 13 train drivers accepted the money, while the remaining 20 drivers wanted to participate for free out of personal interest or support for the research project. Participants provided informed consent and all procedures followed the principles of the Declaration of Helsinki.

## 3.3 Apparatus and stimuli

*3.3.1 Technical setup*

Interviews were conducted in online video meetings via Zoom and the sessions were video-recorded. Due to technical problems, four interviews had to be conducted via phone call. In two cases, the images were presented via Zoom (as only the audio did not work) and in two cases they were emailed to the respective participant. The phone calls were audio-recorded.

*3.3.2 Instruction video*

An instruction video explained the procedure of the study. Participants were informed that they would see 15 images and would be asked five questions about each image (see below for details). To answer the questions about counterfactuals (i.e., what would make the situation less or more dangerous), participants were explicitly told that they could imagine all kinds of changes to the situation, and were provided with examples for such changes (e.g., object size and distance, type of object, environmental conditions). The video took 2:20 minutes, relied on an animated Microsoft PowerPoint presentation, and was presented in German.

*3.3.3 Images*

Participants saw static images taken from the perspective of the train and showing situations in which collisions might occur (see Figure 1 for examples). We aimed to create a diverse stimulus set that is suitable to elicit a variety of relevant factors. Accordingly, the selection of images was based on a pilot study in which we had interviewed six train drivers about their experiences with obstacles and collisions. Based on these interviews, we gathered an initial selection of images and tested it with an experienced train driver. Based on his comments, we discarded some images and added others, leading to a final selection of fifteen images. A description of their contents is provided in Table 1.

The images varied on several dimensions. First, the potential obstacles in focus were either people (8 images), objects (4 images), or both (3 images). People varied in several features such as their identity (e.g., pedestrians, children, track workers), proximity to the tracks (e.g., on, next to, far away), and actions (e.g., waiting for the train, leaving the tracks, playing). Objects additionally showed considerable variation in their physical features (e.g., size, mass). Furthermore, the images varied in numerous context factors such as the type of railway area (e.g., station, shunting yard, branch line, main line) and other aspects of their infrastructure (e.g., curvature, state of the tracks, foliage, fencing) or natural environment (e.g., weather, time of day). In terms of more abstract features, the images varied in their complexity (i.e., number of hazardous elements to consider) and presumed riskiness (i.e., everyday scenes, situations in which collisions were imminent).

**Figure 1.** Example images used during the interviews. (A) Platform, (B) Pram, (C) Track worker, (D) Group.

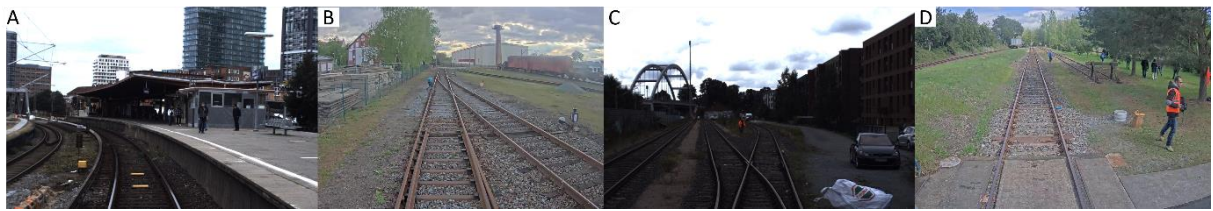

**Table 1.** Image contents with description, type of railway area, and average rating of risk (answer to the question "How dangerous is the situation?" rated on a scale from 1-10).

| No | Image | Description | Area | Rating |
|---|---|---|---|---|
| 1 | Platform | Several people on a platform, all behind the safety line, dark areas under a roof, track workers at a distance | Suburban train station | 1.9 |
| 2 | Person on platform | Single person standing at the end of a platform, staring towards the tracks, wearing a hood | Train station, main line | 2.2 |
| 3 | Pedestrian crossing | Person about to enter the pedestrian level crossing in a station, cyclist approaching, signal at danger | Suburban train station | 4.5 |
| 4 | Level crossing | Unsecured level crossing, cyclist just leaving the tracks, car on the tracks, another car approaching, bus next to the tracks | Urban area, branch line | 6.0 |
| 5 | Person on tracks | Person walking between the rails, towards the train | Shunting yard | 7.9 |
| 6 | Crutches | Elderly person leaving the tracks with crutches, small yellow object on the right rail head | Shunting yard | 6.8 |
| 7 | Child | Child and parent walking next to the tracks in rainy weather, child turned away from the train, wearing a hood | Shunting yard | 7.6 |
| 8 | Group | Group of people behind trees, one person leaving the tracks, one in the foreground turned away from the train, wearing a safety vest and carrying a camera | Shunting yard | 4.9 |
| 9 | Track worker | Track worker walking between two tracks behind a switch, big bag next to the tracks | Urban area, branch line | 4.4 |
| 10 | Construction site | Railway construction site with three workers, one behind a fence, two next to the tracks without a fence, excavator on the left | Narrow gauge railway | 5.1 |
| 11 | Sheep | Five sheep next to the tracks, track worker and locomotive on the tracks at a distance | Rural area, branch line | 6.7 |
| 12 | Pram | Pram on the tracks, turned away from the train driver | Shunting yard | 8.0 |
| 13 | Cardboard | Flat pile of cardboard boxes on top of a switch | Shunting yard | 3.8 |
| 14 | Leaning tree | Tree leaning towards the tracks | Wooded area, main line | 3.7 |
| 15 | Fallen tree | Birch on the tracks in a forest | Wooded area, branch line | 8.1 |

## 3.4 Procedure

An overview of the procedure is provided in Figure 2. The interviews were conducted by the first author (RM), while the second author (JS) took notes. At the start of an interview, demographic information was collected unless train drivers had already participated in the pilot study. Subsequently, they watched the instruction video and had the opportunity to ask questions.

During the interview, participants saw 15 images in randomised order. For each image, they had to answer five questions: (1) How dangerous is the situation? (2) Why? (3) What would make the situation less dangerous? (4) What would make the situation more dangerous? and (5) How ambiguous is the situation?

**Figure 2.** Procedure of the study.

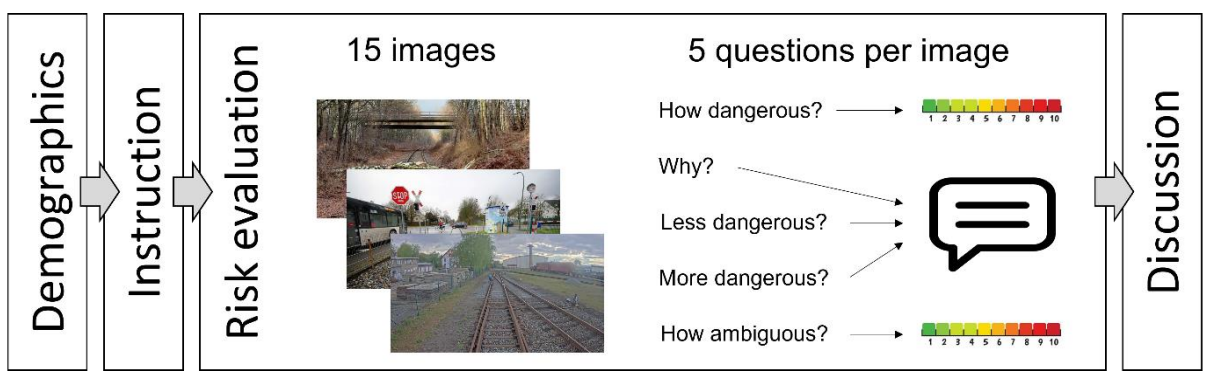

The two ratings for questions 1 and 5 were provided on 10-point Likert scales ranging from "very little" to "very much". These ratings only served for an approximate comparison of the images but were not analysed. For collision risk, a precise evaluation is closely tied to operational factors (e.g., train speed). As these factors cannot be decoded from static images, an image could receive any rating from 1-10, depending on participants' assumptions about these factors. For ambiguity, it turned out that some participants evaluated whether they knew what was happening, some evaluated whether they knew how to react, and many of them changed the aspect they rated between images. As these ratings often pointed in opposite directions, the numbers are not interpretable. While answering questions 1 and 2 (i.e., how dangerous is the situation and why?), participants sometimes indicated that the situation already was as harmless or as dangerous as it can get. In these cases, the interviewer left out question 3 or 4 about factors making the situation less or more dangerous, respectively. Moreover, before moving to the next question, the interviewer asked participants whether they wanted to add any factors, unless they had already reported several factors on their own. Other than that, the interviewer only asked clarification questions.

The risk evaluation phase of the interview (i.e., from the first to the last image) took between 28 and 90 minutes ($M$ = 54.4, $SD$ = 14.5). After completing all 15 images, participants were asked whether any additional factors affected their risk evaluation that had not been covered in the interview so far. Moreover, they were debriefed and had the opportunity for further discussion with the researchers.

### 3.5 Data analysis

All interviews were transcribed according to the five questions asked for each image. We excluded information that was not relevant to the study (e.g., comments about politics, private information) and to ensure anonymity, we cleared the transcripts of all information that might identify the participant (e.g., names of locations). The transcripts were then used to code the concepts that train drivers used to evaluate collision risk. To this end, all risk-relevant statements were highlighted in the transcripts. These statements could be factual or counterfactual (i.e., present or absent in the situation). They included observable cues (e.g., person is close to the tracks, person does not wear safety vest, weather could be rainy) as well as non-observable concepts like people's mental states (e.g., person cannot hear the train). Statements about the train driver's mental states were coded but excluded from the data analysis (e.g., I would see the person in time, I do not know what the person is doing). Statements about the train driver's actions were not coded (e.g., I would operate the whistle, I would brake).

After train drivers' statements were highlighted in the transcripts, they were transferred to a tabular format (one table per image). Each column represented a train driver and each row represented a concept. A concept was defined as a prototypical statement subsuming all statements with identical content but minor differences in formulation. This resulted in 237 to 574 distinct concepts per image (not including concepts about train drivers' mental states). However, it is somewhat subjective whether to code two statements as referring to the same or two different concepts. Therefore, the unit of analyses will not be individual concepts but clusters of concepts.

To derive these clusters, the concepts were structured into factors, subfactors and fine-grained categories. This structure was not predefined but emerged from the data in a bottom-up manner based on a qualitative content analysis. On the highest level, eight factors were extracted. To describe potential obstacles, statements were assigned to the factors *People* or *Objects & Animals*. To describe the context, statements were assigned to the factors *Overall Impression*, *Environment*, *Infrastructure*, *Train Operation*, or *Safety Measures*. Finally, statements regarding the *Consequences* of collisions were collected. These eight factors were further divided into subfactors (e.g., People was divided into *Identity*, *Location*, *Movement*, *Action*, *Physical Features*, and *Mental States*). An overview is provided in Table 2. A further subdivision into fine-grained categories was performed during coding. However, this procedure led to 167 categories in total, and reporting the results on this level of detail is beyond the scope of the present article (the data are made available via the Open Science Framework). An exception will be made for the two obstacle factors (i.e., People, Objects & Animals), given their prominence in train drivers' risk evaluations. Thus, the six subfactors used to describe obstacles were further split into fine-grained categories (e.g., Mental States was split into *Emotions & Sanity*, *Attentional Focus*, *Perception*, *Thoughts & Expectations*, *Knowledge*, *Intentions*). An overview is provided in Table 3.

To analyse what inferences train drivers drew from the cues they considered, the contents of the coding tables were transferred to the graphics editor yEd. Concepts were coded as labelled boxes and the relations between them were coded as lines connecting these boxes. Given the large number of concepts per image, the resulting relation graphs are not suitable for gaining an overview. However, by clicking on a concept, one may inspect which other concepts it has been linked to. The graphs are made available via the Open Science Framework. In the present article, we will only provide a qualitative description of how train drivers generated interpretations and anticipations from prior knowledge, context factors, indicator objects, and the potential obstacle.

**Table 2.** Factors and subfactors used to describe potential obstacles, their context, and the consequences of collision.

|  | Factors | Subfactors |
|---|---|---|
| Obstacle | People | Identity, Location, Movement, Action, Physical Features, Mental States |
|  | Objects & Animals | Identity, Location, Movement, Action, Physical Features, Mental States |
| Context | Overall Impression | Dynamics, Multiple Influencing Factors, Mood, Structuredness |
|  | Environment | Light & Sight, Weather, Noise, Territory, Other |
|  | Train Operation | Speed, Driving Characteristics, Vehicle Characteristics, Other |
|  | Infrastructure | Type, Design, State, Other |
|  | Safety Measures | Constructive Safety Measures, Signals & Warning, Personal Surveillance, Other |
| Collision | Consequences | Obstacle, Vehicle, Train Driver, Passengers & Freight, Infrastructure, Other |

**Table 3.** Splitting the subfactors into fine-grained categories for the two obstacle types (People and Objects & Animals).

| Subfactors | Fine-grained categories |
|---|---|
| Identity | Presence & Type, Number |
| Location | Proximity to Train, Relation to Tracks, Relation to Other Objects, Other |
| Movement | Movement Type, Movement Direction |
| Action | General Actions, Interaction with People & Objects, Communication with Train Driver, Mishaps, Rule Violations |
| Physical Features | Clothes & Gadgets, Health & Functioning, Posture & Rotation, Bodily Features |
| Mental States | Emotions & Sanity, Attentional Focus, Perception, Thoughts & Expectations, Knowledge, Intentions |

# 4 Results

## 4.1 Statements and context factors

In sum, 11.741 statements were coded, with an average of 23.7 statements per train driver and image. These numbers strongly varied between drivers, ranging from 3-68 statements per image ($SD$ = 11.9, for averages see Figure 3A). However, train drivers' average number of statements did not correlate with their job experience, $r$ = -.028, $p$ = .877 (see Figure 3B). Statements were assigned to eight factors (see Figure 3C). Not surprisingly, most statements referred to the two types of obstacles, *People* (39.8 %) and *Objects & Animals* (18.5 %). The contents of these statements will be reported in Section 4.2.

**Figure 3.** Numbers of statements per train driver and factor. (A) Average number of statements per image for each driver, sorted by number of statements. Each bar represents one driver. (B) Relation between statements and job experience. Each dot represents one driver. (C) Shares of statements for each of the eight factors.

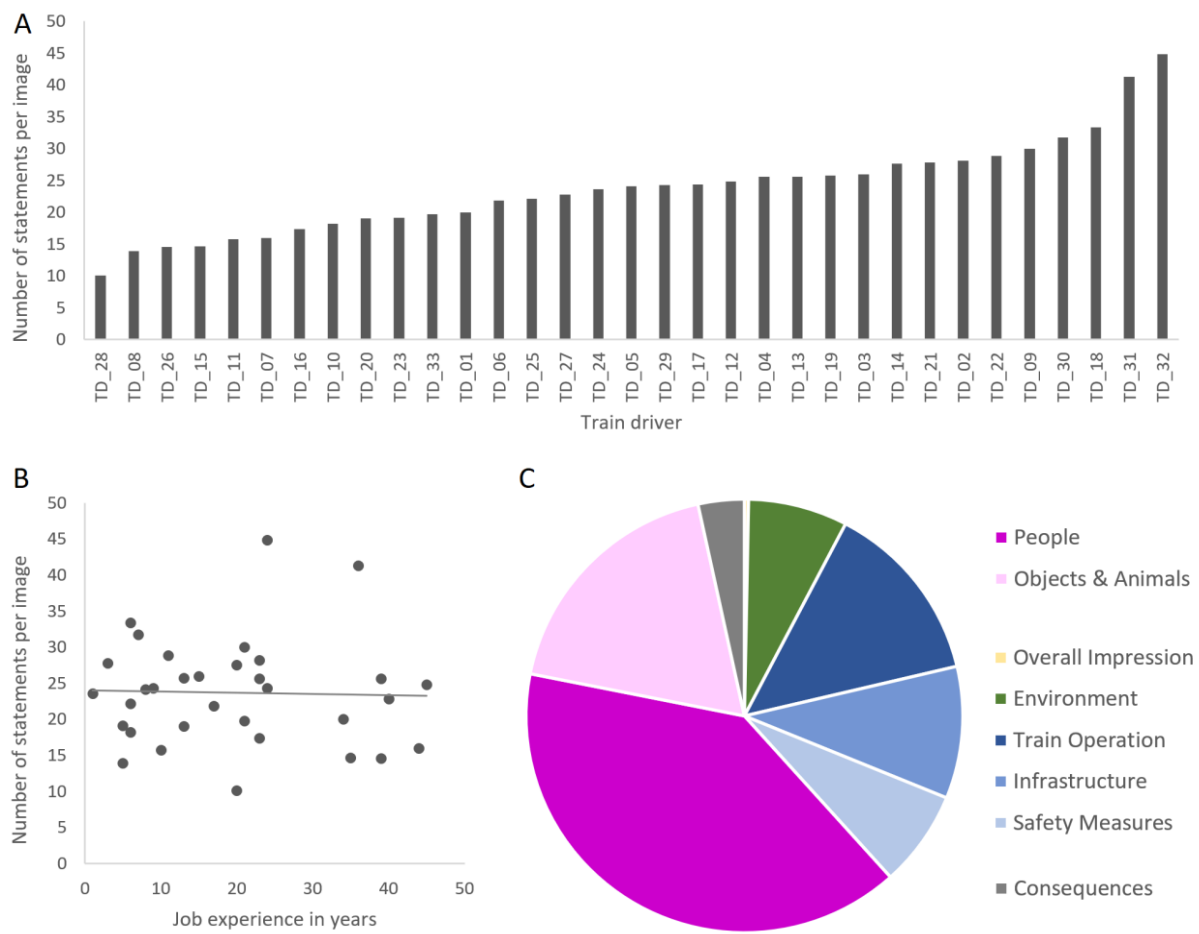

A substantial number of statements tied collision risk to context factors (see Figure 4A). Holistic descriptions of a situation's risk potential in the form of *Overall Impressions* were rare (0.3 %). Instead, train drivers heavily relied on specific cues to evaluate collision risk. First, they considered the *Environment* (7.4 %). The vast majority of these statements addressed light and sight conditions (e.g., sun is not shining, it could be darker, object is occluded). The weather was relevant as well (e.g., it is rainy, there could be a storm), while little emphasis was placed on the territory (e.g., it is a wooded area, there could be a nursing home nearby).

Most statements about context tackled issues of *Train Operation* (13.6 %). Here, train drivers predominantly considered the speed of the train, arguing that situations are less dangerous when driving more slowly. This focus on speed was strongest for images where people actually came in contact with the tracks, either on level crossings or while trespassing. Other frequent operational characteristics concerned the mode of operation (e.g., lower risk for shunting), the ability to brake (e.g., increased braking distances), as well as signals and their aspects (e.g., signal at danger). Fewer statements were made about train characteristics (e.g., freight train, train is heavy) and other features of train operation (e.g., advance information, prior knowledge, other trains).

Another context factor that received considerable attention was *Infrastructure* (9.9 %). Here, train drivers related their risk evaluations to the type of railway area (e.g., station, shunting yard, open track) and how it was designed (e.g., platform is narrow, no overhead lines, site is located behind a curve). Moreover, the state of the railway area was considered a risk factor, with two key issues being foliage (e.g., pruning is insufficient) and the state of the tracks (e.g., rails are rusty).

Regarding *Safety Measures* (7.2 %), train drivers mostly discussed safety by design (e.g., barriers are absent, fence does not keep people from entering the area), whereas fewer statements concerned signals and warnings (e.g., announcements at stations, St. Andrew's Cross) or personal surveillance (e.g., track protection officer, someone should guide pedestrians across the tracks).

Finally, train drivers contemplated the *Consequences* of collisions (3.4 %). Interestingly, they mostly did this when the obstacles were objects or animals, but rarely when they were people (see Figure 4B). Only three images with people had an enhanced frequency of consequence-related statements (i.e., Crutches, Track worker, Sheep). However, these images included both a person and a highly salient object or animals, and references to consequences were almost completely restricted to the latter.

**Figure 4.** Frequency of statements about context and consequences. (A) Statements for different factors and subfactors describing the context of a potential collision. (B) Statements about possible consequences of collisions, plotted for each image. Bars are blue when people were present in the image, orange when they were wearing a safety vest, and grey when there were no people. One grey bar has a darker shade, because a person was present far in the background.

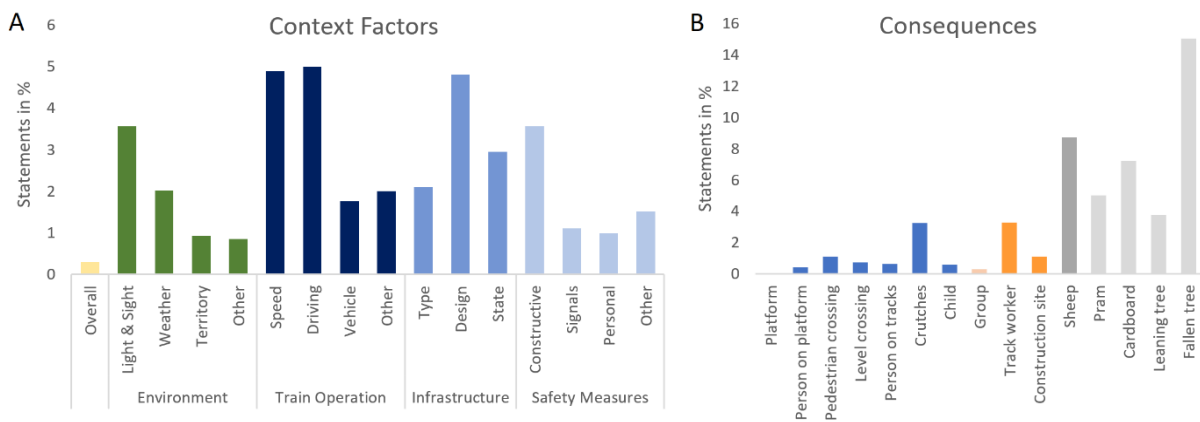

## 4.2 Reasoning about obstacles

*4.2.1 Comparing people and objects: Identity, Location, and Physical Features*

The subfactors train drivers used to reason about the two obstacle types are presented in Figure 5. When discussing people, most statements referred to their *Identity* (9.6 %). Moreover, train drivers relied on other observable cues like *Location* (6.9 %), *Movement* (5.9 %), and *Physical Features* (4.9 %). In addition to these observable cues, they reasoned about people's abstract *Actions* (5.2 %), thus assigning meaning to what these people are doing. Interestingly, *Mental States* (7.1 %) were used more often than any other subfactor besides Identity. Apparently, interpreting what might go on inside people's heads seemed almost as relevant as who they are, and more relevant than where they are, what they do, or what they look like. In contrast, when train drivers discussed the risk associated with objects or animals, statements about *Identity* (7.8 %) clearly dominated all other subfactors. Thus, train drivers seemed most concerned with what an object is, rather than contemplating its detailed features (see Figure 5A).

**Figure 5.** Comparing statements about people versus objects & animals. (A) Identity versus all other subfactors. Fine-grained categories for (B) Location and (C) Physical Features.

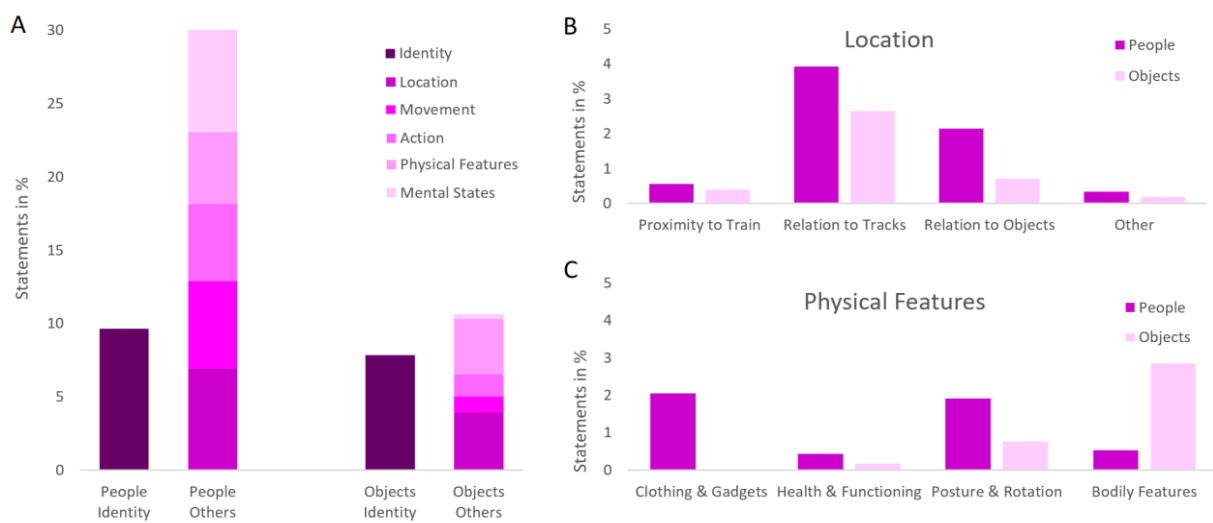

Two subfactors can easily be compared between the two obstacle types. In case of *Location*, the same pattern emerged for people and objects (see Figure 5B): train drivers rarely mentioned the obstacle's proximity to the train but rather focused on its relation to the tracks (e.g., child is close to the tracks, tree could already be on the tracks) or its relation to other objects (e.g., person is behind the safety line, tree could hang in the overhead lines). Conversely, train drivers described people and objects quite differently in terms of their *Physical Features* (see Figure 5B). The overall number of statements was similar (4.9 % for people, 3.8 % for objects & animals), but the specific contents differed considerably. For people, many statements were about *Clothing & Gadgets* (e.g., worker is wearing a safety vest, person could wear headphones). Unsurprisingly, this category was not used for objects and animals. However, an interesting double dissociation was observed for situation-specific versus permanent physical features. On the one hand, train drivers used situation-specific features of *Posture & Rotation* much more often when reasoning about people than objects (e.g., worker is turning his back towards the train, person could be sitting). On the other hand, they used permanent *Bodily Features* extensively to describe objects but not to describe people (e.g., tree is small, box could be heavy). For the remaining subfactors used to reason about obstacles (i.e., Movement, Action, Mental States), it makes little sense to compare them between people and objects. This is because they are of little relevance to describe static objects, and dynamic objects like cars being operated by people were coded as people. Therefore, in the following two sections, we will only focus on people.

*4.2.2 What are people doing: Movement and Action*

When reasoning about the risk associated with people, train drivers paid close attention to what they were doing. In fact, when adding all statements about Movement (5.9 %) and Action (5.2 %), their joint share was higher than that of Identity (9.6 %), suggesting that identification is insufficient to evaluate the risk of colliding with people. However, the usage of these behavioural concepts strongly depended on the specific image (see Figure 6).

*Movement* was mentioned most often when a person was on the tracks and thus the risk of collision was imminent unless the person would move. In this case, movement made up 13.3 % of train drivers' total statements. Movement also was used a lot when people moved away from the tracks or when several people moved in different directions. Generally, movement direction relative to the tracks was more important than the type of movement (e.g., running, not standing still).

Movement is a rather straight-forward concept that leaves little room for interpretation. In contrast, understanding people's actions requires train drivers to reason about their behaviour on a higher level of abstraction. Such inferences strongly depended on image content. First, people's *General Actions* (e.g., person might wait for the train) was mentioned most often for an image showing a lonely person on a platform and this person's intentions were somewhat ambiguous. Conversely, general actions were mentioned least often for two images showing level crossings. This tendency to avoid abstract behavioural descriptions for level crossings was also observed for all other action categories, except for rule violations (see below). The second action category, *Interaction with People & Objects*, stood out for two images where it was used much more often than for all other images: a child playing with her mother next to the tracks and a pram left on the tracks. While considering interactions seems obvious in the former case, it was somewhat surprising in the latter, given that no people were present. Still, train drivers interpreted why people might have left the pram and anticipated how they might try to save it. Another type of interaction is *Communication with the Train Driver*. This category was highly relevant to reason about track workers but not about ordinary people. The fourth action category, *Mishaps*, was used most often for an impaired person leaving the tracks with crutches, for groups of people on a platform, and for workers being too close to the tracks. Finally, train drivers reasoned about *Rule Violations*. Ironically, this category was used least often for a person walking on the tracks, because the coding scheme assigned this to the subfactors Location and Movement. The image with the most frequent statements about rule violations was the one with the person using crutches. Here, train drivers reasoned about the person stepping on the rail head and having placed an object on the tracks. Placing objects also resulted in frequent statements about rule violations for the two images showing a pram and a cardboard box on the tracks. In both cases, the objects triggered train drivers to reason about people's actions although no people were visible.

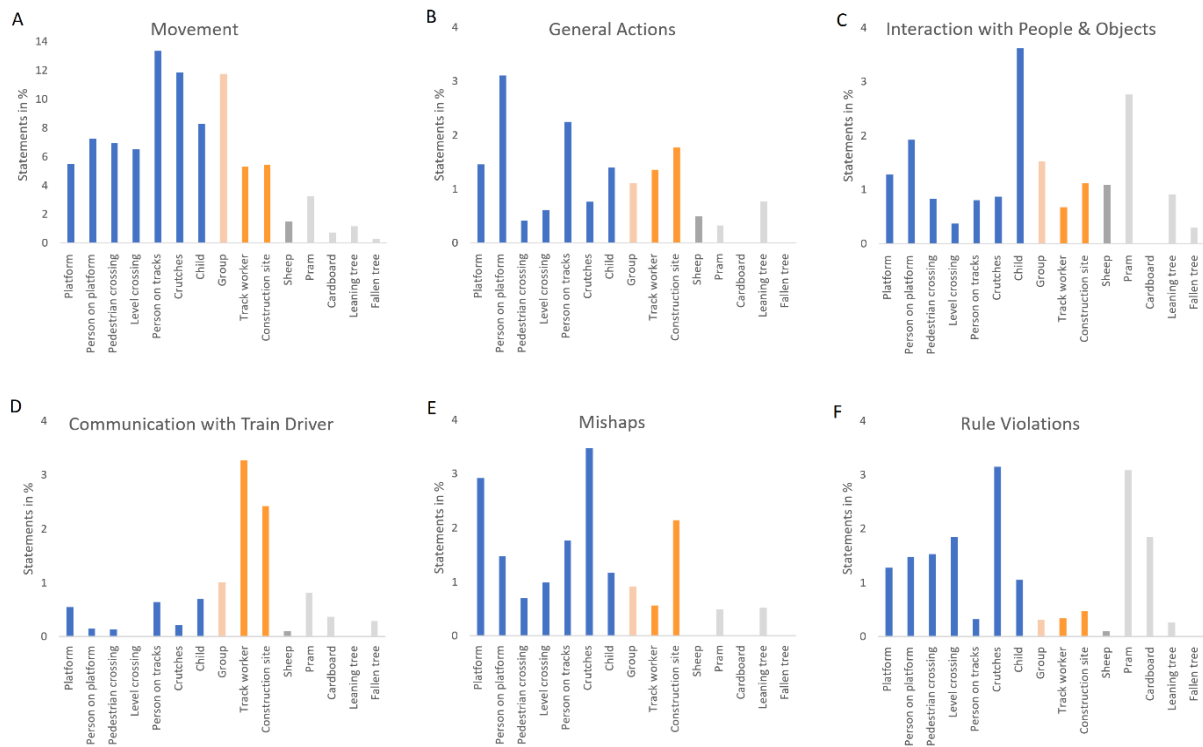

**Figure 6.** Frequency of statements about movement and action for individual images. (A) Movement, (B) General Actions, (C) Interaction with People & Objects, (D) Communication with Train Driver, (E) Mishaps, and (F) Rule Violations.

*4.2.3 What might go on inside people's heads: Mental States*

While train drivers heavily relied on mental states to reason about people, the specific contents of mentalising varied between images. Overall, these statements were rare when no people were present in the images (see grey bars in Figure 7), but there also were variations between images showing people (see blue and orange bars). First, *Emotions & Sanity* were rarely considered when people wore safety vests. When encountering track workers, train drivers apparently had other concerns than whether they are in control of themselves. Moreover, these statements were rare for people inside vehicles. Interestingly, one image with no person visible triggered train drivers to use this category: a pram abandoned on the tracks. Statements about people's *Attentional Focus* were rare overall, except for three images with people turning their backs towards the train. Conversely, train drivers frequently reasoned about people's *Perception*. This category was used most often for a child with her mother and for a person on the tracks. These were the only two images showing people in acute danger, and in both cases, it was ambiguous whether these people were able to perceive the train. Surprisingly, train drivers rarely mentioned perception for an elderly person leaving the tracks with crutches. However, they considered it highly unlikely for this person to return, which apparently made it seem less relevant what she can or cannot perceive. For *Thoughts & Expectations*, the distribution was rather even and no systematic dependencies on image content became apparent. This was different for a related concept: statements about *Knowledge* were predominantly used for track workers, but rarely for ordinary people. Generally, safety vests made train drivers reason about people's knowledge. Accordingly, such statements were also used when the person wearing the vest was a photographer. Moreover, train drivers frequently reasoned about people's *Intentions.* Here, three images stood out. For one, intentions were rarely mentioned for the two images showing track workers. Their intentions seemed to be as non-salient to train drivers as their emotions and sanity. Conversely, the image that triggered most statements about intentions was the abandoned pram, although no person was visible.

**Figure 7.** Frequency of statements about mental states for individual images. (A) Emotions & Sanity, (B) Attentional Focus, (C) Perception, (D) Thoughts & Expectations, (E) Knowledge, and (F) Intentions.

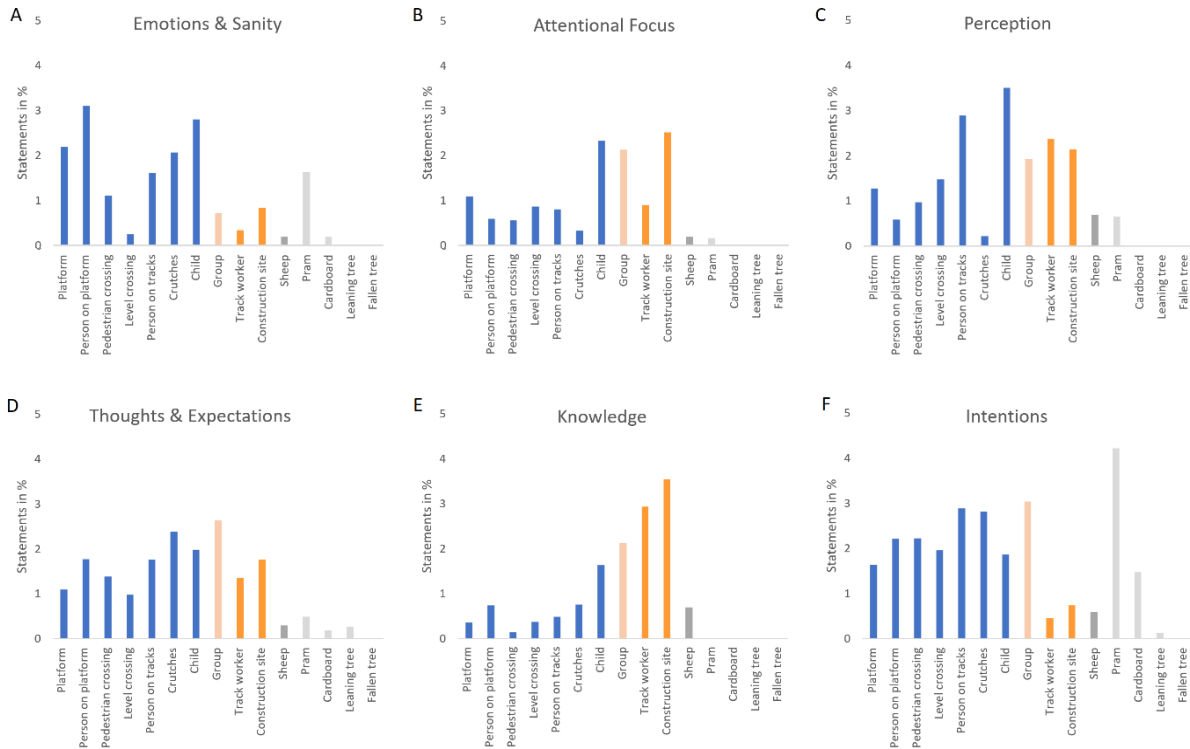

## 4.3 Connecting concepts to interpret and anticipate events

So far, train drivers' statements were assigned to separate categories and their distribution was analysed. However, understanding how train drivers reason about the risk of collision requires us to go one step further, asking how they interpret and anticipate events. That is, how do they connect concepts from different categories to infer what a situation could mean and how it might develop? An example is provided in Figure 8. The following sections will show how train drivers used their prior knowledge, context factors, indicator objects, and features of the potential obstacle to interpret and anticipate the presence of obstacles as well as the behaviour and mental states of people.

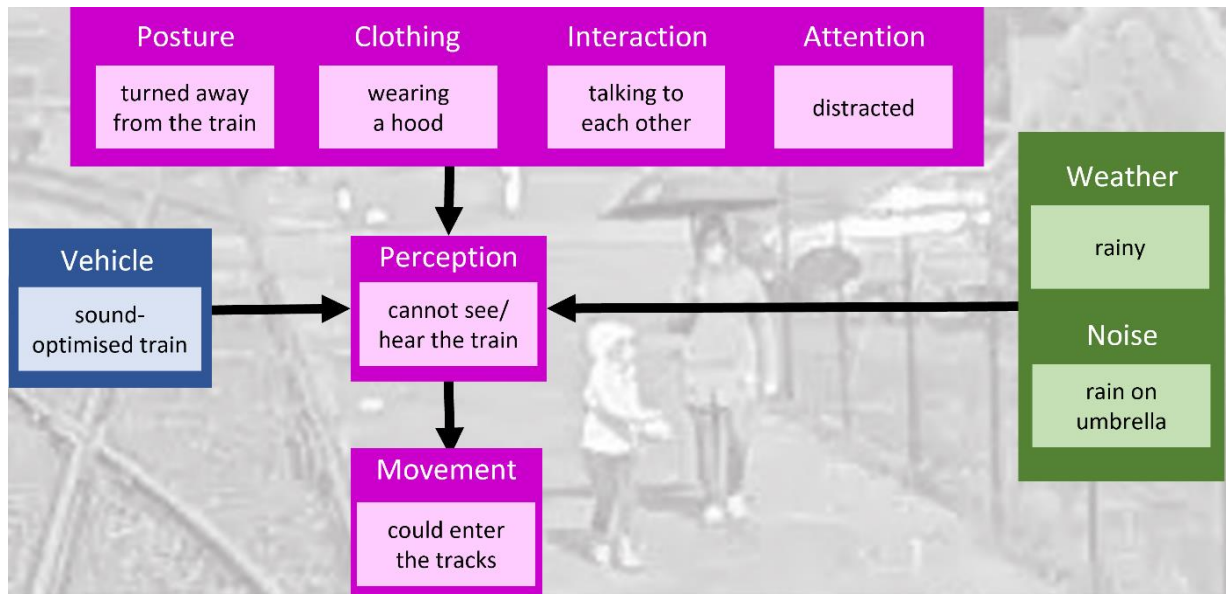

**Figure 8.** Train drivers used a variety of cues to infer mental states, and then used these states to anticipate behaviour. Colour marks indicate concepts about people (purple), train operation (blue), and environment (green).

*4.3.1 Prior knowledge*

Train drivers rely on their *route knowledge* to anticipate obstacles. For instance, in case of a tree leaning towards the tracks, several drivers commented that in reality, they would know about its history. If the tree had been leaning like this for months, they would not be worried at all. Knowing one's route also means being aware of nearby places that attract people. For instance, one train driver explained how knowing about a particular lake allowed him to expect people crossing the tracks during summer season. Similarly, train drivers expect people to appear on known illegal paths crossing the tracks or in the vicinity of residential buildings. However, not only train drivers' anticipation but also their risk evaluation depends on route knowledge. For instance, one driver argued that he would be highly alarmed if a person appeared in the woods, unless he knew that a hiking trail ran alongside the tracks, which was frequently used at weekends. Aside from physical characteristics, route knowledge also encompasses familiarity with train schedules that can be used to infer what people might be doing or intending. In case of a lonely person on a platform, one train driver said that usually, he would know whether a passenger train is about to arrive. In this case, a person standing there would be normal. Otherwise, this might increase his suspicion that the person might intend to commit suicide.

Besides route knowledge, train drivers use *other sources of knowledge* to anticipate the presence and behaviour of obstacles. For instance, knowing about a recent storm allows them to expect fallen trees. Knowing about public events does not only allow them to expect large groups of people at stations, it also enables inferences about their mental states and behaviour: people coming home from public events are more likely to be drunk and might act in an uncontrolled or aggressive manner. Finally, train drivers use their general knowledge to infer people's mental states (i.e., Theory of Mind). For one, they contemplated what people can or cannot know: people approaching a level crossing do not know whether the train's signal is set to danger, and workers do not know how the switches are set. Moreover, train drivers have preconceptions about specific types of people, for instance when expecting trainspotters to be focused on their historical train and thus less likely to attend to the opposite direction.

*4.3.2 Context factors*

Train drivers anticipate risky events based on contextual cues. First, environmental factors such as time of day allow them to anticipate the *presence and identity of obstacles*. At night, deer is more likely to cross the tracks, whereas people are less likely to be outside. Train drivers also rely on counterfactuals, generating inferences from the absence of particular cues. For instance, when trees are not swaying in the wind, fallen trees are unlikely to obstruct the tracks. When there is no fence, people are more likely to trespass. Second, context factors such as time or infrastructure allow train drivers to anticipate the *behaviour of people*. At daytime, fewer people are putting themselves at risk. At level crossings or illegal paths crossing the tracks, people's behaviour is more predictable. Even operational factors are used to anticipate people's behaviour. One train driver stated that people at level crossings were more likely to wait for a short passenger train to pass than for a long freight train. Third, context is used to infer *people's mental states*. In this regard, the weather was mentioned particularly often. Rain prevents people from seeing and hearing the train, while making them eager to get to a dry area, thus strongly narrowing their focus. Similarly, infrastructure predicts people's perceptual abilities, for instance when dense foliage at level crossings obstructs their view of the tracks. Infrastructure can also provide cues to people's intentions. For instance, train drivers reasoned that people may indeed intend to harm themselves or their child, but probably not in a shunting yard.

*4.3.3 Indicator objects*

Train drivers draw inferences based on objects. First, objects and people can predict the *presence of other objects and people*. For instance, one fallen tree or one child predicts more to come. However, objects do not only predict other objects of the same type, but can sometimes predict more dangerous objects. For instance, the presence of railway detonators led one driver to anticipate a defective train blocking the tracks behind the next curve. Similarly, objects can predict people. The most striking instance was an image showing a pram, which led several train drivers to infer that the mother was likely to show up. One train driver said: "Where there's a pram, there's a mother. That's a very simple basic law. And the mother will come. And the mother will come even more quickly and be even more distraught when I am arriving with my locomotive." Second, objects can predict *people's behaviour*. For instance, one train driver stated that a bike can make people less mobile, because people will not easily let go of it when it gets stuck in the tracks at a level crossing. Other drivers argued that people are often incapable of handling their heavy baggage at stations, increasing their risk to enter the danger zone. Finally, objects are used to infer *people's mental states*. For instance, crutches suggested to train drivers that an elderly person might be confused, and motor scythes suggested that a worker might not hear the train. A gravel stone on the rail head suggested that children might be intending to see the train run over it, therefore hiding nearby and perhaps getting hit by stone splinters. Even the presence of people is used to infer other people's mental states, such as when the presence of track protection officers indicates that workers are aware of the train, or the presence of children indicates that parents might be distracted.

*4.3.4 Potential obstacle*

Finally, train drivers interpret features of potential obstacles to make inferences about them. First, they use people's behaviour and appearance to reason about *their future behaviour*. When people are pacing, swaying, or fooling around in a group, train drivers infer that someone might enter the tracks.

Conversely, sitting decreases people's mobility and thus their risk of collision. Relevant behaviour also included explicit communication with the train driver. Most importantly, if workers are raising their hand, train drivers know they will stay away from the tracks. Second, train drivers use people's behaviour and appearance to infer *their mental states*. If people were talking to each other or turning their backs towards the train, drivers considered them less likely to perceive the train. Conversely, gazing towards the train was seen as a major factor to decrease risk. Among the physical features used to infer mental states, clothing in general and safety vests in particular seemed highly informative. Safety vests made drivers conclude that people are knowledgeable and want to be seen. Not wearing the vest in the prescribed manner led them to speculate that a worker might not be taking safety very seriously. Even general clothing is used to predict mental states. Dirty or ill-fitting clothes indicates to train drivers that people might be under the influence of alcohol. Wearing a hood can make them conclude that a person might not perceive the train. Wearing a hood in bright sunshine can make them suspect suicidal intentions. Conversely, wearing summerly clothing suggests that people are more likely to be on their way to a lake than intending to end their lives. None of these cues is perfectly reliable, and train drivers sometimes disagreed about the inferences derived from a particular cue. However, when combining all these different cues in a highly situation-specific manner, this allows train drivers to go far beyond what is currently visible.

# 5 Discussion

In the future, AI-based systems might contribute to avoiding collisions with trains, but their current focus on object identification seems overly limited. To reason about requirements for AI, it is important to understand how human train drivers evaluate the risk of collision. Therefore, we conducted image-based expert interviews with experienced train drivers. They rated how dangerous each situation was, explained why, and discussed which factors would decrease or increase the risk. We extracted the concepts train drivers used to reason about collisions. This approach complements previous behavioural studies of train drivers' eye movements and object detection (e.g., Du et al., 2022; Ma et al., 2024; Sun et al., 2019). While these studies can reveal what train drivers look at, we considered it essential to understand what train drivers actually see and how they use this information to interpret events and anticipate future developments. Such issues have partly been addressed in previous analyses of how train drivers navigate their visual world (e.g., Karvonen et al., 2011; Naweed & Balakrishnan, 2014; Rosić et al., 2022; Roth & Multer, 2009). However, we aimed to go beyond anecdotal evidence and systematically collect the concepts train drivers use to generate situation awareness. In the following sections, we will highlight some key findings, discuss the resulting challenges for AI, make some limitations of the present study explicit, and conclude with an outlook for future research.

## 5.1 Summary of key findings

As suggested by previous studies (Karvonen et al., 2011; Rosić et al., 2022; Roth & Multer, 2009), train drivers did not merely identify potential obstacles to evaluate the risk of collision. In this regard, a first key finding is that *train drivers heavily rely on context*. This is in line with a previous conclusion that landmarks like houses and bridges serve as cues to what is likely to happen (Roth & Multer, 2009). In the present study, train drivers used the type of railway area and its proximity to residential buildings or public attractions to anticipate people's presence, behaviour, and mental states. Similarly, their risk evaluation incorporated the temporal and environmental context such as time of day, weather conditions, or characteristics of the trains and tracks. These are valid cues that can indeed predict the likelihood and severity of collisions (Zhang et al., 2023). Thus, context enabled train drivers to interpret ambiguous situations and anticipate what is likely to happen.

A second key finding relates to the question of how far one can get with object identification. Our findings suggest that it depends on the obstacle: *train drivers use different information to reason about people and objects*. For objects, their identity accounted for a large share of the statements, and the remaining statements mainly referred to observable cues like position, movement, and size (cf. Rosić et al., 2022). In principle, these cues can be made available to an AI system. Thus, to predict the risk of colliding with objects, relying on observable cues might get you quite far. This is not the case for people, where train drivers actively inferred concepts that were not directly observable, such as the meaning of people's actions and their mental states. Many of these "soft factors" are valid predictors of collisions, such as people's pre-collision behaviour or alcohol use (Zhang et al., 2023). Thus, going beyond simple identification pays off. Another difference between people and objects relates to the consequences of collision. For objects, train drivers evaluated danger as a combination of likelihood and severity, while for people, they evaluated risk in a stricter sense, only referring to likelihood but not severity. This could either have strategic reasons as collisions with people are distressful (Limosin et al., 2006), or it might simply be less informative to ponder about consequences in case of people, because collisions almost always kill the person but almost never harm the train (Hampel et al., 2023; Rosić et al., 2022).

A third key finding is that *train drivers reason about people's mental states*. They infer them from people's body language and appearance, which corroborates previous anecdotal evidence (Karvonen et al., 2011; Rosić et al., 2022). However, they also infer them from prior knowledge, context factors, indicator objects, and a number of other cues relating to the people in question. In fact, mental states were the second-most frequent category used to reason about people. Although this has rarely been considered in the context of train driving, it resonates with ample empirical evidence on mentalising: people constantly consider what might go on inside other people's heads (Frith & Frith, 2005). Thus, are our findings trivial? Perhaps not, as they can help specify what types of inferences train drivers make in what situations. For instance, we observed that they avidly reason about knowledge but not emotions or intentions in case of track workers, while the opposite was found for ordinary people. Similarly, train drivers inferred a lack of attention when people were turning their backs towards the train. Such systematic relations between observable cues and inferences about mental states can help disambiguate situations and potentially be used in technical systems.

### 5.2 What does this mean for AI?

Overall, the present findings suggest that it is necessary to go beyond simple object identification when aiming to avoid collisions. But what are their implications for AI? Applying our findings in technical systems seems challenging for at least two reasons: train drivers' reasoning cannot easily be translated into rules and it cannot fully be made available in training data.

First, one might conclude that the concepts and relations obtained in studies with train drivers could simply be fed into algorithms. However, a problem with this approach is that *all simple rules fail*. Let us illustrate this with a few examples. A simple rule might be that potential obstacles are more dangerous when they are located on the tracks or in the danger zone than when they are far away. However, in case of a level crossing, several train drivers were not concerned about a car on the tracks, because they assumed that once they arrive, the car would be gone. Instead, they were worried about a car that still was several meters away from the tracks but approaching (cf. Rosić et al., 2022). Another rule might be that small objects are less dangerous. However, when a small yellow object was placed on the rail head, some train drivers concluded it was a brake shoe that might make the train derail. Moreover, one train driver thought it was a railway detonator announcing a defective train. Thus, small objects can cause or predict major problems (cf. Rosić et al., 2022; Tichon, 2007). Yet another rule might be that situations are less dangerous when no people are to be seen. However, an abandoned pram suggested to train drivers that a distraught person might appear to save the pram. Well, but at least more people should be more dangerous, shouldn't they? While this was indeed what many train drivers said, the image considered least dangerous was the one with the highest number of people: a platform of a suburban train station. Under some circumstance, people are normal. Moreover, in case of a lonely person on a platform, some train drivers argued that the situation would be less dangerous if more people were present, because people usually do not commit suicide in groups. Given that no simple rule seems satisfactory, what about more complex combinations of rules? Using the previous example, one might assume that a single person is less dangerous than a group unless no train is scheduled to stop at a station, unless the person is standing at the end of the platform, unless she is wearing a hood in bright sunshine... the combinations are getting absurd rather quickly, and still cannot capture the complexity of the real world. No pre-conceived combination of rules will cover all relevant instances.

Thus, an alternative approach is to use AI systems that do not rely on explicit rules but learn to evaluate situations via training. This is what most automatic train operation endeavours are aiming for. But could this basic principle be extended beyond object identification to capture the complex relations that train drivers are using? A problem with this approach is that it is *unclear how to provide suitable training data*. First, train drivers rely on context, and most of this context is not available to technical systems. While it would be easy to provide some context information (e.g., train speed, infrastructure), other information is harder to get (e.g., there is a hiking path nearby, at this season people leave the path to gather mushrooms). A second problem is that AI systems need suitable training data (e.g., images or videos), and the available data may be systematically biased. While dataset bias is a general problem for AI (Torralba & Efros, 2011), there are additional difficulties specific to the railway domain. For instance, images are often generated on branch lines as it is problematic to place obstacles on the open tracks. Thus, the AI might learn the wrong contexts and therefore not work robustly under real-life conditions (cf. Beery et al., 2018). While this problem seems solvable, there is another, more fundamental problem of providing suitable training data: the concepts and relations used by train drivers cannot be captured in images or videos. For instance, how to express that a person cannot see the train, or that wearing a hood makes it less likely to see the train? Train drivers base these inferences on their everyday interactions with the world, which are not available to AI systems.

So how to move forward? Is there any way in which the present findings can inspire AI research? We hope that they can provide a vision of what is desirable and make it explicit where there are still gaps between contemporary AI systems and the cognitive activities of train drivers. While we have focused on collision avoidance, similar gaps have been highlighted for other aspects of train driving (Jansson, Olsson, et al., 2023; Karvonen et al., 2011; Tripathi & Borrion, 2016). Some might conclude that these gaps cannot be bridged and thus train drivers are impossible to replace. Others might see it as a challenge to improve their AI systems. In any case, researchers, decision makers, and the general public should be aware of the gaps. This might lead us to rethink what kind of human-machine systems we are aiming for, and what kinds of technologies we need to realise them.

### 5.3 Limitations of the present study

We need to critically assess the reliability and validity of our findings, as a number of methodological and conceptual limitations may constrain what we can and cannot infer from them. A first, obvious limitation is that we used *static images instead of dynamic scenarios*. This was a purposeful choice in order to avoid the pitfalls of implicit perception and evaluation: by introducing an obstacle (i.e., lack of dynamic cues), we hoped to get train drivers to make it explicit which of these cues are missing to evaluate the risk of collision. Our results suggest that this approach worked, as train drivers repeatedly explained which dynamic cues they would need. On the other hand, it might be that without these cues, they did not even think of certain aspects of the situations that would matter to them in real life.

Second, a limitation regarding data analysis concerns our approach of *counting train drivers' statements*: how informative are the numbers we reported? One aspect of this question is the reliability of the numbers as they strongly depend on image contents. If we had used fewer images with people, train drivers would have made fewer statements about people. If we had used other images with other people, the shares of categories would have differed, as suggested by the variations between images shown in Figures 4, 6 and 7. In short, one should take a key finding of our study to heart: context matters. Future studies should systematically vary image contents to specify how this affects the results. Another aspect of counting statements is the validity of the numbers. Frequency is not the same as relevance, and there can be different reasons for making a statement. Train drivers might name a particular concept because they want to disambiguate the referent (e.g., describe a person's location to make sure the interviewer knows which person is meant), because the concept is easy to use (e.g., salient features like safety lines on platforms make it easier to specify the location of people), because it is easy to imagine that it might change (e.g., a person could be gone vs. a bridge could be gone), or because it is not the focus of the image (e.g., a person and locomotive in the background, while there are sheep in the foreground). Conversely, and somewhat paradoxically, train drivers might not mention concepts that are relevant in every situation and thus do not differentiate (e.g., most train drivers chose not to talk about train speed or sight conditions for every image). Thus, naming or not naming concepts can have other reasons than relevance. Follow-up studies should use the concepts elicited in the present study and directly assess how train drivers rate their relevance for evaluating the risk of collision.

A third methodological limitation is that *we cannot quantify inferences*. We argued that situation awareness depends on interpretation and anticipation, but our interview method does not allow us to determine exactly how often train drivers did that. For instance, statements like "the person could walk towards the tracks" might be anticipations based on observable cues, or they might simply be responses to our instruction to explain how the situation could be different. Still, it is interesting what kind of changes train drivers made. They did not merely report observable counterfactuals like identity, location, or movement, although the method would have allowed for that. Instead, they actively inferred concepts and specified the relations between them. Thus, although we cannot quantify these inferences, the qualitative analysis presented in Section 4.3 show that they were made on a regular basis.

A fourth limitation concerns the difference between verbal reports and actually *being in the situation*. We tried to make our interviews as concrete as possible by using images as anchors, but this obviously differs from actually driving a train. First, we neither asked nor coded how train drivers would react to an obstacle. While this has been described elsewhere (Rosić et al., 2022), we asked how train drivers evaluate potential collisions rather than avoid them. However, actions and action opportunities are a major influence on how people perceive and evaluate situations (Gibson, 1979; Neisser, 1976). A related limitation is that we did not investigate to what extent train drivers actually use the cues they reported. On the one hand, they might not have reported all the cues they are actually using. This could be due to cognitive limitations in verbal reasoning, the ability to imagine counterfactuals, or other traits. Accordingly, train drivers might not be fully aware of their own information requirements and mental processes. On the other hand, train drivers might not actually use all the cues they reported. Attending to them might be desirable but not always be feasible in practice, due to high train speed, working conditions, or cognitive limitations. For instance, high workload decreases train drivers' perception and anticipation (Zoer et al., 2014), and problems of distraction, inattention, or fatigue contribute to many rail accidents (Baysari et al., 2009; Naweed, 2013). Given that we cannot know for sure whether train drivers reported the cues they use and use the cues they reported, the implications for train driving remain unclear. Would collisions actually be reduced when using these cues? Some of the cues are known to be valid indicators of collision risk and severity (Zhang et al., 2023). However, for most of the more indirect, inferred concepts such as people's mental states, cue validity remains an open issue for future research.

### 5.4 Outlook and future research

There are many exciting prospects for future research. One is to study *interindividual differences*. Our train drivers did not only differ in their number of statements, but also in their emphasis of particular concepts. For instance, some focused on operational factors and others on the environment. Some focused on people's actions and others on mental states. Train drivers also differed in the breadth of their focus, with some only considering their own situation and others including the risk for other trains. Such differences did not result from job experience simply expressed in the number of years. However, the experience with particular collision scenarios might matter. Train drivers with collision experience search the environment more and therefore are better at detecting obstacles (Ma et al., 2024). Moreover, cognitive abilities such as verbal reasoning, fluid intelligence, and selective attention can affect driving performance (Collins, 2024). Future studies should assess what factors affect train drivers' cue usage and inferences.

Moreover, our findings could inspire research on *train driver training*. In principle, collections of cues and inferences could be incorporated into training scenarios (Tichon, 2007). Novice drivers could learn how to recognise relevant cues, how to interpret them, and how to use them to anticipate future developments. This could help them develop more suitable mental schemata, moving away from stereotypical procedures and towards more adaptive, situation-specific thinking. However, this raises the question of what can be learned via instruction, given that expertise rests on extensive experience with particular situations (Klein, 1989, 2008). Therefore, future research should investigate how knowledge about relevant cues and inferences can be incorporated into training activities to make experiences more transparent and to help novice drivers sharpen their focus.

Finally, future research should consider how risk evaluation is affected by *modern technologies.* Such technologies can harm train drivers' cognitive performance (Naghiyev et al., 2016; Naweed, 2014): attending to new in-cab interfaces conflicts with monitoring the environment, and technologies like ERTMS/ETCS can reduce anticipation and impair decision-making. This is in line with a long research tradition revealing how automation can negatively affect human performance (Bainbridge, 1983; Parasuraman & Manzey, 2010; Parasuraman & Wickens, 2008). Thus, unless train drivers are fully removed from the cab, it should be investigated how the envisioned AI-based technological support would impact their ability to establish situation awareness. A challenge for future research is to design an effective cooperation between train drivers and AI.

## 5.5 Conclusion

When train drivers evaluate the risk of collision, they engage in complex cognitive activities that go far beyond object identification. If AI systems are supposed to emulate these activities or at least generate similar outcomes, this might require other kinds of technologies than the ones currently envisioned. The present study cannot provide simple prescriptions on how to develop AI. Instead, we hope that it may support AI researchers and practitioners in asking the right questions and developing realistic visions.

# Acknowledgments

We are grateful to all train drivers who participated in this study and provided fascinating insights into their risk evaluation processes. Moreover, we want to thank the Union of German Train Drivers (GDL) for support in participant acquisition, Carsten Knoll for support in preparing the transcripts, all participants of a "train drivers meet AI" workshop for inspiring discussions about the results of this study, as well as Rustam Tagiew and Kristin Mühl for providing valuable feedback on the manuscript. This work was supported by the German Centre for Rail Traffic Research (DZSF) at the Federal Railway Authority within the project "Explainable AI for Railway Safety Evaluations (XRAISE)".